\documentclass[acus]{JAC2003}

\usepackage{epsf}
\usepackage{color}
\usepackage{graphicx}

\begin{document}

\newcommand{\ccdot}{ \! \cdot \! }
\newcommand{\Br}{{\cal B}}
\newcommand{\beq}{\begin{equation}}
\newcommand{\eeq}{\end{equation}}
\newcommand{\la}{\langle}
\newcommand{\cO}{{\cal O}}
\newcommand{\ra}{\rangle}
\newcommand{\beqa}{\begin{eqnarray}}
\newcommand{\eeqa}{\end{eqnarray}}
\newcommand{\dis}{\displaystyle}
\newcommand{\ba}{\begin{array}}
\newcommand{\ea}{\end{array}}
\newcommand{\no}{\nonumber}
\newcommand{\yuk}{{Y}}
\def\Green{} % PANTONE 323
\definecolor{mygreen}{rgb}{0,0.6,0}
\newcommand{\green}[1]{\textcolor{mygreen}{#1}}
\newcommand{\blue}[1]{\textcolor{blue}{#1}}
\newcommand{\red}[1]{\textcolor{red}{#1}}
\def\Black{}
% --------------- abbreviated journal names -------------------------

\def\ap#1#2#3{     {\it Ann. Phys.  }{\bf #1} (#2) #3}

\def\arnps#1#2#3{  {\it Annu. Rev. Nucl. Part. Sci. }{\bf #1} (#2) #3}
\def\npb#1#2#3{    {\it Nucl. Phys. }{\bf B #1} (#2) #3}
\def\plb#1#2#3{    {\it Phys. Lett. }{\bf B #1} (#2) #3}
\def\pr#1#2#3{     {\it Phys. Rev. }{\bf   #1} (#2) #3}
\def\prd#1#2#3{    {\it Phys. Rev. }{\bf D #1} (#2) #3}
\def\prb#1#2#3{    {\it Phys. Rev. }{\bf B #1} (#2) #3}
\def\prep#1#2#3{   {\it Phys. Rep. }{\bf #1} (#2) #3}
\def\prl#1#2#3{    {\it Phys. Rev. Lett. }{\bf #1} (#2) #3}
\def\ptp#1#2#3{    {\it Prog. Theor. Phys. }{\bf #1} (#2) #3}
\def\rpp#1#2#3{    {\it Rept. Prog. Phys. }{\bf #1} (#2) #3}
\def\ppnp#1#2#3{   {\it Prog. Part. Nucl. Phys. }{\bf #1} (#2) #3}
\def\rmp#1#2#3{    {\it Rev. Mod. Phys. }{\bf #1} (#2) #3}
\def\zpc#1#2#3{    {\it Z. Phys. }{\bf C #1} (#2) #3}
\def\ijmpa#1#2#3{  {\it Int. J. Mod. Phys. }{\bf A #1} (#2) #3}
\def\sjnp#1#2#3{   {\it Sov. J. Nucl. Phys. }{\bf #1} (#2) #3}
\def\yf#1#2#3{     {\it Yad. Fiz. }{\bf #1} (#2) #3}
\def\jetpl#1#2#3{  {\it JETP Lett. }{\bf #1} (#2) #3}
\def\epjc#1#2#3{   {\it Eur. Phys. J. }{\bf C #1} (#2) #3}
\def\ibid#1#2#3{   {\it ibid. }{\bf #1} (#2) #3}
\def\jhep#1#2#3{   {\it JHEP  }{\bf #1} (#2) #3} 
\def\physica#1#2#3{{\it Physica }{\bf A #1} (#2) #3}
\def\ncim#1#2#3{   {\it Nuovo Cim. }{\bf #1} (#2) #3}

%\hfill  INFNNA-IV-2003/29
\renewcommand{\thefootnote}{\fnsymbol{footnote}}

\title{CP, CPT and Rare Decays
%\centerline{\large
 \thanks{Invited talk at Alghero DA$\Phi$NE Workshop, 10-13 September 2003}}

\author{Giancarlo D'Ambrosio,  INFN-Sezione di Napoli, 80126 Napoli ITALY\thanks{giancarlo.dambrosio@na.infn.it}}

\maketitle

\begin{abstract}
We describe possible kaon physics goals to achieve with a high luminosity $\Phi-$factory.
We motivate the relevance to improve the present bounds on CPT kaon physics quantities. 
Also, the interferometer machine   $\Phi-$factory is useful to study CP violating/conserving
effects in kaon decays. Thus we investigate   $K \to 3\pi $ amplitudes:  charge asymmetries are 
interesting in charged kaon decays. In the case of neutral kaon decays one can study  
directely $K_S \to 3\pi $ or interferences effects. Interference may also be used to study
the CP even  $K_S \to \pi^+ \pi ^-\pi^0 $ and final state interactions.
% also in light 
%of recent interesting theoretical and experimental results.
%and  $K \to \pi  \pi \gamma $
%decays.
\end{abstract}
\section{Introduction}

Recentely the majestic CPT tests \cite{Hagiwara:fs,Maiani:ue,Thomson:pm,D'Ambrosio:1994wx} 
in kaon physics have been challenged by neutrino physics
\cite{Mura03} where it has been argued that neutrino physics probes a  shorter scale
than the  typical scale probed by 
kaon physics. I recall, as I will show later,   that kaon physics maybe  
sensitive to Planck scale physics \cite{Buchanan:1991ce,Ellis:1992dz}.
We take the pragmatic approach that  an high luminosity $\Phi-$factory
must improve the CPT kaon physics tests to match the  neutrino physics level.
Bell-Steinberger relations, dictated by the  unitarity conditions, are the main tool  
to improve the CPT kaon 
physics  bounds. 
Now these bounds are limited by the $K\to \pi \pi$ measurements, so it is compulsary to improve 
these  experimental results.
We will review the CPT neutrino bounds first, then  the present bounds in kaon physics from 
Bell-Steinberger relations. Finally 
we mention other interesting non-CPT violating physics issues achievable
at an  high luminosity $\Phi-$factory.  

\section{CPT violation}
Relativistic quantum field theories predict  a very important 
property: CPT invariance \cite{Greenberg:2003nv}, which holds under the following three hypotheses:
{\begin{itemize}
\item Lorentz invariance
\item Hermiticity of the Hamiltonian
\item Locality.
\end{itemize}}
CTP-violation and/or the accuracy on which we test CPT is fundamental in physics   
and searched  in several experiments \cite{Bluhm:2003ne}. 
 A thoretical acceptable framework to generate CPT-violation is the one suggested
in Ref.   
\cite{CGK}, where  departures from Lorentz invariance  generate     
CPT-even and  CPT-odd terms: small non-invariant  terms are added to the Standard
Model Lagrangian, these are assumed renormalizable (dimension $\le \ 4$ ), invariant
 under $SU(3)_c\times SU(2)_L\times U(1)_Y$ and rotationally and translationally invariant in a preferred frame 
(then fixed to be the one where the cosmic microwave background is isotropic). 
String theory is presumably valid up  to the Planck scale and argued to be  CPT-conserving. But spontaneous
 CPT-violation is still allowed: S-matrix elements may violate CPT according to the details of the low
energy limit.
In fact  string theory can generate Lorentz and CPT violating terms \cite{Kostelecky:1989nt}.
 Actually there are also explicit
 quantum field theory examples of spontaneous
 CPT-violation \cite{Jenkins}.
Just to give an explicit example of Lorentz violation we mention the one particularly used in cosmic rays
and neutrino tests. We change the coefficient of the square of the magnetic field in the Lagrangian of quantum 
electrodynamics:

\beq
\vec{B} ^2 \rightarrow (1+\epsilon)\vec{B} ^2
\eeq
This will  cause the velocity of light  $c $, given by $c^2=1+\epsilon$   to differ 
from the maximum velocity of particles, which remains equal to one.
This is just one of the  terms to be  added to the Standard
Model Lagrangian \cite{CGK}, invoking an explicit violation of Lorentz invariance.

Recentely Ref.\cite{Barenboim:2002tz} has challenged CPT by giving up locality but
 not Lorentz invariance; they add to the usual Dirac term  the following non-local fermionic action 
\[
\mathbf{S}=\displaystyle{{\frac{i\eta }{\pi
}}\int d^{3}x\int dtdt^{\prime
}\,\bar{\psi}(t,\mathbf{x})\,{\frac{1}{t-t^{\prime }}}\,\psi
(t^{\prime },\mathbf{x}).}
\]
In order to prove that a  \raisebox{0.14cm}{\rotatebox{300}{\Bigg|}}${\hbox{\hskip-25pt CPT}}$ 
 lagrangian can generate physical amplitudes
it is important to check causality: 
however it is still disputed if this model 
produces a satisfactory CPT-violating model \cite{Greenberg:2003ks}.

\vskip0.6cm
\centerline{\large \bf \raisebox{0.14cm}{\rotatebox{300}{\Bigg|}}${\hbox{\hskip-25pt CPT}}$  in $\nu$'s 
and challenge} 
\vskip0.3cm
%{\large \bf ($\delta \equiv \Delta m^2_\nu -\Delta m^2_{\bar{\nu}}$)}
%\section{$CPT { \large \hskip-20pt \Bigl/}\quad$  in $\nu$'s and challenge 
%($\delta \equiv \Delta m^2_\nu -\Delta m^2_{\bar{\nu}}$) }
As we shall see in the kaon sector CPT and quantum mechanics are already tested to an interesting level \cite{Hagiwara:fs}
\beq
|m_{\bar{K}} -m_K|<  10^{-18} m_K\Longrightarrow
|m_{\bar{K}}^2 -m_K^2|< {0.5}\ eV^2.
\eeq
This already probes an interesting size: in fact quantum gravity may generate a
CPT violating kaon mass term of order \cite{Buchanan:1991ce}:
\beq
\frac{m_K^2}{M_P} \quad .
\eeq
Murayama   \cite{Mura03} has wondered if neutrino can challenge this limit.
In fact neutrino   sector plays now  an  important role in   flavour studies,
 and new experiments in their attempts to pin down the neutrino flavour matrix, 
the PMNS matrix analogous to 
the CKM matrix, 
will give us    also useful information  for  CP and CPT sudies.
 
Already the present limits on   solar neutrino and antineutrino mass  difference:  
\beq
|\delta|\equiv |\Delta m ^2 _\nu -\Delta m ^2 _{\bar{\nu}}|   
<{1.3 \times 10^{-3}} \  eV^2 \, (90\% C.L.)\label{eq:mnu}
\eeq
constitutes an interesting challenge for kaon physics. Though of course, there is no
reason {\it a priori} to expect the same amount of CPT violation in the two systems.

This  scale is comparable to the measured  dark energy
\beq
\rho _\Lambda \sim {(2 \times 10^{-3} eV)}^4 
\eeq

We remind also that  different 
spectra for neutrinos and antineutrinos has been invoked in order explain 
 LSND data \cite{Murayama:2000hm}.

\centerline{\large \bf \raisebox{0.14cm}{\rotatebox{300}{\Bigg|}}${\hbox{\hskip-25pt CPT}}$  in the $K$'s
 mass and width matrix }
%\section{$CPT {\hskip-20pt \Bigl/}$\hskip1cm in the $K$'s
% mass matrix}
\vskip0.3cm

Though there are arguments \cite{QMV}  suggesting possible quantum mechanics violations and possible tests
in interferometry machines like $\Phi$-factories and CPLEAR \cite{huet}, here we assume that  
conservation of probability has  a stronger validity than CPT, thus we can
keep unitarity but relax CPT violation. In the kaon system 
we 
can  describe mass and decay eigenstates  
by the diagonalization \cite{Maiani:ue,D'Ambrosio:1994wx,mannelli}
 
$$
\left(\begin{array}{cc}
M_{11} - i\Gamma_{11}/2 &
M_{12} - i\Gamma_{12}/2 \\ & \\
M_{21} - i\Gamma_{21}/2  &
M_{22} - i\Gamma_{22}/2
\end{array}\nonumber\right)$$

$$ {\rm CPT} \Longrightarrow M_{11}=M_{22} \quad \Gamma_{11}=\Gamma_{22},$$
with the eigenvectors
    $$K_{S,L}  = 
       {  \dis\frac{ \left[ \left(1 + \epsilon_{S,L}\right)K^0 +
                  \left( 1 - \epsilon_{S,L} \right) \bar{K}^0\right] }
{\sqrt{2\left( 1 + |\epsilon_{S,L}|^2\right)}}}$$
Encoding in $\Delta$ the CPT violating contributions
  \begin{eqnarray}
\Delta&=&\frac{\frac{1}{2} \left[{ M_{11} - M_{22} -\frac{i}{2}
    \left( \Gamma_{11} -\Gamma_{22}\right)} \right]}
{m_L - m_S +i(\Gamma_S - \Gamma_L)/2}\label{Delta}
\end{eqnarray}
we can write 
\begin{eqnarray}
\epsilon_{S,L} & = &\frac{
    -i \Im\left( M_{12}\right) -
            \frac{1}{2} \Im\left(\Gamma_{12}\right)
                    }{
    m_L - m_S +i(\Gamma_S - \Gamma_L)/2
                    }\mp { \Delta}  \nonumber\\
& =& \epsilon_M \mp { \Delta} \nonumber 
\end{eqnarray}
\beq
\epsilon_M\equiv |\epsilon_M| e^{i{\varphi _{SW}}}\qquad 
\tan {\varphi _{SW}}=\frac{\dis{2(m_L -m_S)}}{\dis{\Gamma_S-\Gamma_L}}. 
\label{phiSW}
\eeq
Thus unitarity predicts the phase of mass CP violation in terms of $\Delta m$
\cite{Hagiwara:fs,KteV}
\beq
\varphi _{SW}=(43.46\pm 0.05)^0\label{eq:.phiew}
\eeq
\vskip0.2cm
{\large \it  \raisebox{0.14cm}{\rotatebox{300}{\Bigg|}}${\hbox{\hskip-25pt CPT}}$ in semileptonic decays}
\vskip0.1cm

 %\subsection{$CPT {\hskip-20pt \Bigl/}$\hskip1cm
% in semileptonic decays}
We will discuss the semileptonic decays of neutral kaons without assuming the $ \Delta S = \Delta Q$ rule
and the CPT symmetry \cite{Maiani:ue,D'Ambrosio:1994wx,mannelli}. \par
The $ \Delta S = \Delta Q$ rule is well supported by experimental
data and is naturally accounted for by the Standard Model, where the $ \Delta
 S = - \Delta Q$ transitions are possible only with two  effective
weak vertices. Explicit
calculations in the SM give a suppression factor of about $10^{-6}$--$10^{-7}$
\cite{guberina}.
Furthermore in any quark model, $ \Delta S = - \Delta Q $
transitions can be induced only by operators with dimension
higher than 6 and therefore should be suppressed \cite{Maiani:ue,Franzini:2004bi}.
However large 
violation of
the $ \Delta S = \Delta Q$ rule does not conflict with any general
principle. We can write \cite{Maiani:ue,D'Ambrosio:1994wx}
\begin{eqnarray}
   A(K^0 \to l^+\nu \pi^-) &=& a + {b} \nonumber \\
   A(K^0 \to l^-\nu \pi^+) &=& c +{d}  \nonumber \\
  A(\bar{K}^0 \to l^-\nu \pi^+) &=& a^* -{b^*} \nonumber \\
  A(\bar{K}^0 \to l^+\nu \pi^-) &=& c^* -{d^*}\nonumber
\end{eqnarray}
CPT implies $ b=d=0$, CP implies $\Im(a)=\Im(c)=\Re(b)=\Re(d)=0$,
T requires real amplitudes and $ \Delta S = \Delta Q$ implies $c=d=0$. Then

$$\delta_{S,L} = \frac{  \Gamma_{S,L} ^{l^+} - \Gamma_{S,L} ^{l^-}}{ \Gamma_{S,L} ^{l^+}+
 \Gamma_{S,L}^{l^-}} =  2\Re(\epsilon_{S,L}) + 2 \Re\left({b
\over a}\right) \mp 2\Re\left({d^* \over a}\right)$$

\begin{equation}
\delta_S-\delta_L\propto {\Re{\Delta},\Re{d^*}}
\end{equation}
Thus a  non-vanishing value of the
difference $\delta_S -\delta_L$ would be an evidence of
CPT violation, either in the
mass matrix or in the ${\Delta S=-\Delta Q}$ amplitudes ($ \Delta$ and
$ d^*/a $ cannot be disentangled by semileptonic decays alone).
The sum $\delta_S +\delta_L$ has CPT-conserving ($\Re(\epsilon_M)$) and
CPT-violating ($\Re (b/a)$) contributions that cannot be disentangled.
\vskip0.6cm
{\large \it \raisebox{0.14cm}{\rotatebox{300}{\Bigg|}}${\hbox{\hskip-25pt CPT}}$ in $K\to \pi \pi$}

%\subsection{$CPT {\hskip-18pt \Biggl/}$\hskip1cm
% in $K\to \pi \pi$}
\vskip0.3cm
CPT can be violated in the mass according to eq.(\ref{Delta}) and in the amplitudes.
If CPT is not conserved   in $K\to \pi \pi$ the new amplitudes ${B_I}$'s appear:  
\begin{eqnarray}
   A(K^0 \to \pi  \pi (I)) &\equiv & (A_I + {B_I})e^{i\delta_I} \nonumber \\
   A(\bar{K^0} \to \pi  \pi (I) ) &
\equiv & (A_I^* - {B_I^*})e^{i\delta_I} \nonumber\end{eqnarray} 
Defining as usual 
$$ \eta_{+-}= {\displaystyle{\frac{A(K_L \to \pi^+  \pi ^-)}{A(K_S \to \pi^+  \pi ^-)}}}
\quad
\eta_{00}= {\displaystyle{\frac{A(K_L \to \pi^0  \pi ^0)}{A(K_S \to \pi^0  \pi ^0)}}}$$
and noticing that (\ref{eq:.phiew}) is approximately equal, in the CPT limit, to the phase of $\epsilon '$,
then the $\eta$'s phases must be equal in the CPT limit. In fact the following CPT bound has been established
experimentally
\cite{Hagiwara:fs,mannelli,KteV,Takeuchi:2002sr} 
%{$B_I$}  is 
%\raisebox{0.14cm}{\rotatebox{300}{\Bigg|}}${\hbox{\hskip-25pt CPT}}$
%$CPT {\hskip-16pt \Bigl/}$ 
%\quad as \hfill{($\scriptstyle 
%\eta_{+-}= |\eta_{+-}|e^{i{\phi _{+-}}}\quad \eta_{00}=|\eta_{00}|e^{i
%{\phi _{00}}}$ )}

$${\phi _{+-}}-{\phi _{00}}=0.22\pm 0.45$$
%\hskip4cm{{\rm {\scriptscriptstyle KTEV,NA48}}} 

%\section{Bell-Steinberger relation and $CPT \red{\hskip-30pt \Bigl/}$}
%\section{Bell-Steinberger relation and  }\rotatebox{330}{ ${\hbox{\kern-6pt CPT}}$}
\vskip0.2cm
\noindent
{\large \bf BELL-STEINBERGER  RELATION  AND } 
\centerline{\large \bf \raisebox{0.14cm}{\rotatebox{300}{\Bigg|}}${\hbox{\hskip-25pt CPT}}$ }
\vskip0.3cm
%\rotatebox{310}{\Bigg|}${\hbox{\hskip-16pt CPT}}$ 
%${\hbox{\kern-6pt CPT}}$
If we think that the  probability conservation is valid up to shorter distances than
CPT then even if CPT is violated we can impose tht unitarity must be valid 
\cite{Maiani:ue,Thomson:pm,Bell:mn}.
Then if  we  consider the time evolution of an initial kaon state which is a a quantum 
superposition of
$K_S,K_L$:
 \beq
|{K(t)}\rangle={a_S}|{K_S}\rangle+{a_L }|{K_L}\rangle
\eeq
and we impose probability conservation,  for any $a_S,a_L$, as
\beqa
-\frac{\dis{d}}{\dis{dt}} |\langle{K(0)} |{K(0)}\rangle|^2=&\nonumber\\
&\hskip-1.5cm\sum_f |{a_S}A(K_S \to f)+{a_L} A(K_L \to f)|^2\  .\nonumber 
\eeqa
This turns in a relation among $K_{S,L}$ masses and  widths defined in (\ref{Delta}),(\ref{phiSW})
and all  $K_{S,L}$ branching ratios:
\beq
\Longrightarrow (1+i\tan {\varphi _{SW}})\left[\Re({\epsilon _M})-i\Im({\Delta})\right]=
\sum_f {\alpha _f}\label{eq:imDelta}
\eeq
where we have encoded in $\alpha _f$'s  all the $K_{S,L}$ branching ratios, $ B_{f}^{S,L}$:
$$
{\alpha _f}= B_{+-}^S\eta_{+-},\  B_{00}^S\eta_{00},\  B_{+-\gamma}^S\eta_{+-\gamma},\ 
 {\frac{\tau_L}{\tau_S}} B_{000}^L
{\eta_{000}},.. $$

Now an accurate experimental knowledge of the various ${\varphi _{SW}},\ {\epsilon _M},\ \alpha _f$'s 
($ {\alpha _{\pi\pi}},\
{\alpha _{\pi\pi\gamma}},\ {\alpha _{000}}..$) determines a limit  on 
$\Im(\Delta)$ in  eqs.(\ref{Delta}) and (\ref{eq:imDelta}). 
The largest experimental error is  now coming from $ {\alpha _{000}}$
(SM prediction $1.9\cdot 10^{-9}$); the  published result of CPLEAR
\cite{Angelopoulos:1998xa},
$B_{000}^S<1.4\cdot 10^{-5}$,  and the interesting preliminary limit of  NA48/1 \cite{Aug3p}
 with
$B_{000}^S<3\cdot 10^{-7}$ ($90\%$CL) lead respectively to \cite{Angelopoulos:1999nu}, \cite{Aug3p}
%\hfill{\footnotesize{\green{,KTEV,NA48}}}
\begin{itemize}
\item
CPLEAR\hskip1.5cm $\Longrightarrow {\Im(\Delta)}=(2.4\pm 5.0)\times 10^{-5}$
\item
NA48/1\hskip1.5cm $\Longrightarrow {\Im(\Delta)}=(-1.2\pm 3.0)\times 10^{-5}$
%\item Improved , but still not close enough to the neutrino sector
\end{itemize}
 The  KLOE  results in this channel are very promising \cite{KLOEKS3p} $B_{000}^S<2.1\cdot 10^{-7}$, 
and  would improve the NA48/1 results. 

Now  we can  use these results  in eq.(\ref{Delta}): this CPT  limits are WORST than the neutrino limits in 
eq.(\ref{eq:mnu}). To improve we need a more accurate determination of $ B_{+-}^S, B_{00}^S$, 
%which could be obtained 
obtainable at 
future DA$\Phi$NE  \cite{Franzini:2004bi,DiDomenico:2003tz}.

\section{CP aymmetries in $K^+\to 3\pi$  and final state interaction}
\begin{figure}[t]
\includegraphics[width=7cm,height=4cm]{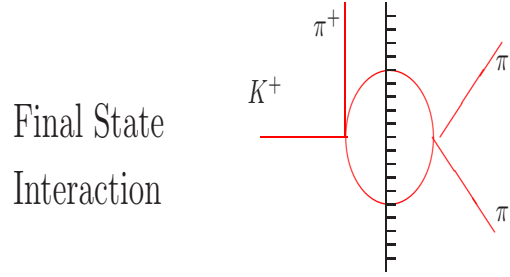} 
 \caption{Final state interaction in $K\to 3\pi$: absorptive contribution.}
\label{fig:phases}
\end{figure}

Direct CP violation in charged kaons is subject of extensive researches at NA48/2 \cite{Sozzi:2003np}. 
Studying 
the $K\rightarrow 3 \pi$ Dalitz distribution in  $Y,X$ \cite{Hagiwara:fs,MP,DI96}
\beq
|A(K\rightarrow 3 \pi)|^2 \sim 1+g\ Y +j\ X + {\cal O}(X^2, Y^2)
\eeq
and determining  both  charged kaon  slopes, $g_{\pm} $, we can define the slope charge asymmetry: 
\beq
\Delta g/2g=(g_+-g_-)/(g_+ + g_-) \label{eq:deltag}.\eeq
There are two independent $I=1$ isospin amplitudes $(a, b)$, 
\beq
%\ba 
A(K^{+} \to \pi^{+}\pi^+\pi^-) = {a} e^{i \alpha_0} + {b} e^{i \beta_0 } Y
+ \cO(Y^2,X^2) 
\label{eq:dg} 
\eeq
with corresponding
final state interaction phases, $\alpha_0$ and $\beta_0$.
 The hope is that $\Delta g$
in  (\ref{eq:deltag}) does {\bf NOT} need to be suppressed by a $\Delta I = 3/2$ transition.
The strong
 phases, generated by the $2 \to 2$ rescattering in Fig. (\ref{fig:phases}), 
are approximated here by their value at the center of the Dalitz plot, but actually have their own
kinematical dependence \cite{IMP} and can be expressed in terms 
of  the Weinberg scattering lenghts, $a_0$ and  $a_2$.
It is particularly interesting to try to write down a  Standard Model (SM)
theoretical expression for $\Delta g/{2g}$,  valid if there is a good chiral expansion 
for the CP conserving/violating $a,b$ amplitudes \cite{DI96,DIP}. In fact under this assumption
we neglect, the $ \cO(p^6)  $ amplitudes $a^{(6)}$, $ b^{(6)}$ and write \cite{DI96,DIP} 
\beqa
\frac{\Delta g}{2g}&=\hskip6.5cm
\nonumber 
%\label{eq:dg}
\\
&\hskip-0.2cm\displaystyle{{\frac{\Im A^0 }{\Re A^0}}}
{{{ (\alpha_0 - \beta_0)}}} \left(\frac{\Re b^{(4)} }{\Re b^{(2)}}-
 \frac{\Im b^{(4)} }{\Im b^{(2)}}+\frac{\Im a^{(4)} }{\Im a^{(2)}}
 -\frac{\Re a^{(4)} }{\Re a^{(2)}}\right)\nonumber\\
&\hskip-3.8cm\sim 22 \epsilon '(\alpha_0 - \beta_0)\sim 10^{-5}.\nonumber 
\eeqa
This result can be improved  by accounting for the kinematical dependence of the strong phases \cite{IMP};
here we have approximated the strong phase difference, $(\alpha_0 - \beta_0)\sim 0.1$ \cite{noi}, 
by its value at the center of the Dalitz plot. More accuracy to determine (\ref{eq:dg}) can be obtained
by evaluating the $\cO(p^6)$ \cite{Gamiz:2003pi}. Also a direct detemination of the strong phases, 
which  is possible through time interferences, as we shall see later,  would be very welcome.
Recentely, a new strategy has been suggested: 
NA48/2 at  CERN \cite{AugIta} 
has 
accumulated a lot of charged kaons, in particular $K^+\rightarrow \pi ^{+} \pi ^{0} \pi ^{0}$, also with 
an accurate scan in
 the  $\pi ^{0} \pi ^{0}$-invariant mass distribution, $M_{\pi ^0\pi ^0}$, finding
a cusp at $M_{\pi ^0\pi ^0}=M_{\pi ^+\pi ^-}$.  This result has been nicely investigated by 
Cabibbo \cite{Cabibbo:2004gq}, which explains the \lq\lq cusp'' 
as an effect 
due  to the opening
of the $\pi ^+\pi ^-$-threshold in Fig. (\ref{fig:phases}). Since the rescattering 
$\pi ^+\pi ^-\to \pi^0\pi^0$ 
is proportional to  $a_0 - a_2$   then 
\beqa
\frac{d\Gamma(K^+\rightarrow \pi ^{+} \pi ^{0} \pi ^{0})}{dM_{\pi ^0\pi ^0}}\bigg| _{\rm NA48} &\Rightarrow 
{\rm cusp \, for}
M_{\pi ^0\pi ^0}=M_{\pi ^+\pi ^-}\nonumber\\
 &\stackrel{\rm cusp}{\Rightarrow} \,
a_0 - a_2.\nonumber\eeqa 
This should give this strong phase difference to a very good accuracy \cite{Cabibbo:2004gq}
once isospin breaking effects are completely under control \cite{Bijnens:2004ku}.
This is particularly exciting due the intense experimental (DIRAC) and theoretical efforts to determine
 $a_0 - a_2$.

%\vskip2cm

%\vspace{4cm}

\section{Time interferences}
One of major advantages at $\Phi$-factories is the known initial   $K_S\ K_L$ quantum state, fixed by 
the  $\Phi$ quantum numbers.
By choosing appropriate final states, $f_1, f_2$, one can study several observables like 
$\Im ( \epsilon'/\epsilon)$ \cite{D'Ambrosio:1994wx}. This has proven to be very difficult so far
\cite{Franzini:2004bi} but there are other   interesting possibilities.

\subsection{$K\to 3 \pi$ and advantages of KLOE/CPLEAR}
Even in the CP limit $K_S$ can decay in $ 3\pi$ if $3\pi$ are in high angular momentum
(and $I=2$) state
$$A(K_S \to \pi^{0}\pi^+\pi^-)={\gamma}{X}
(1+i{\delta _2}) -{\xi}{XY},$$
$\gamma$ and ${\xi}$ are  $\Delta I=3/2$ transitions while
%$X,Y$-Dalitz variables,
  ${\delta _2}$ is the final state interaction phase.
$\gamma$ is predicted by isospin
while ${\xi}$ \cite{Kambor:ah}  and 
${\delta _2}$ by ChPT \cite{noi}.
Since ${\delta _2}$ is very small ($\sim 0.1 $), the branching
$Br(K_S \to \pi^{0}\pi^+\pi^-)$,  is not very sensitive to 
${\delta _2}$.

%\vspace{4cm}
%\begin{figure}[t]
%\includegraphics[width=21cm]{CERN1}  
%\end{figure}
%\end{itemize}

%\begin{itemize}
%\item 

At $\Phi$-factory, choosing as final states, $f=\pi ^+\pi ^-\pi^0,l\pi\nu$, and 
by opportune kinematical cuts   \cite{D'Ambrosio:1992nn} it is possible  to find a time dependent
 asymmetry proportional to
\beq
\displaystyle{ 
\int_+ \Re\left(A^{+-0}_L A^{+-0*}_S \right)}
\left[\cos(\Delta mt)+{\widetilde \delta}\sin(\Delta mt) \right]d\phi_{3\pi},
\label{Rp}
\eeq
where
\begin{equation}
{\widetilde \delta}\simeq\alpha_0 -\delta_{2}\nonumber
\end{equation}
By opportune time dependent studies it is possible to measure
this observable linear in $A^{+-0}_S$ and in the final state phase.
Another interesting channel is  $K_L\rightarrow \pi^+ \pi^-\gamma$, where it is possible to extract, with a high statistics
  $\Phi$-factory,   the direct CP violation component \cite{D'Ambrosio:1994wx,Donoghue:hn,D'Ambrosio:1993yf}.
 Time interferences in $K_L\rightarrow \pi^+ \pi^-e^+ e^-$ may also be interesting \cite{Sehgal:1999vg}.

\section{Conclusions}
At interferometry machines, like  $\Phi$-factories, with high statistics,
let us say $10^{12}$ kaons, the golden searches are $\Im (\epsilon ' /\epsilon)$,
 the semileptonic modes and $\eta _{000}$
and $\eta _{+-0}$\cite{D'Ambrosio:1994wx}. But we think  
that also the CP conserving $A(K_S\rightarrow \pi^+ \pi^-\pi^0)_{CP=+}$ and final state interactions in 
$K\rightarrow 3\pi$ will be very useful. Particularly after the good news from NA48/2 and Cabibbo
  \cite{Cabibbo:2004gq}.
The charge asymmetry limits in $K^\pm\rightarrow 3\pi$ and $K^\pm\rightarrow \pi \pi \gamma  $
\cite{Colangelo:1999kr} are going 
also to be improved and may be tested to an  interesting level.
CPT tests are also a clear target: to this purpose let's stress again that 
we need improvement in $K\rightarrow 2\pi$
amplitudes.
With larger statistics, of course, more is possible, like for instance the interesting time dependent studies  in 
$K_L\rightarrow \pi^0 e^+ e^-$ \cite{timeevol}. The rare kaon decays program is also very rich \cite{ginot}.
%\end{itemize}

\section{Acknowledgments}
I thank Augusto Ceccucci and  Nello Paver  for interesting discussions,
and Luigi Cappiello for collaboration and  stimulating conversation.
Also I thank  G. Isidori for inviting me to this very nice workshop and discussions.
This work  is partially supported by IHP-RTN, 
EC contract No.\ HPRN-CT-2002-00311 (EURIDICE).

\end{document}